# Spectroscopic Studies of two-dimensional Superconductivity

Qiang-Jun Cheng(程强军)[1], Xu-Cun Ma(马旭村)[1,2,3], Qi-Kun Xue(薛其坤)[1,2,3,4] and Can-Li Song(宋灿立)[1,2,3,†]

[1] *Department of Physics and State Key Laboratory of Low-Dimensional Quantum Physics, Tsinghua University, Beijing 100084, China*
[2] *Frontier Science Center for Quantum Information, Beijing 100084, China*
[3] *Beijing Academy of Quantum Information Sciences, Beijing 100193, China*
[4] *Shenzhen Institute for Quantum Science and Engineering and Department of Physics, Southern University of Science and Technology, Shenzhen 518055, China*

Two-dimensional superconductivity has become a major frontier in condensed matter physics. It holds the key to the mechanism of high-temperature superconductors and offers an exceptional arena to stabilize emergent quantum states enabled by enhanced electron correlations in reduced dimensionality. These states are frequently characterized by spatial modulations and intertwined with competing orders, calling for studies that combine real-space imaging with local spectroscopy. Scanning tunneling microscopy and spectroscopy meets this need by directly accessing local density of states with lattice-scale resolution. In this review, we summarize recent advances of the study on several representative unconventional superconductors using this technique, focusing on direct characterization of high-temperature superconducting planes, pair-density waves, and topological superconductivity in both artificial heterostructures and intrinsic materials. We conclude by outlining current challenges and future directions motivated by the microscopic insights.

## 1. Introduction

Superconductivity, characterized by zero electrical resistance and perfect diamagnetism, has captivated condensed matter physics for more than a century.[1-3] For conventional superconductors, the Bardeen-Cooper-Schrieffer (BCS) theory provides a well-established framework in which phonon-mediated, isotropic $s$-wave pairing with spatial homogeneous order parameter dominates (Fig. 1(a)).[4] High-temperature ($T_c$) superconductors, however, revealed a profoundly different paradigm.[5-19] In cuprates, iron-based superconductors (IBSCs), and the recently discovered nickelates, superconductivity is primarily confined within quasi-two-dimensional (2D) structural units—the $CuO_2/NiO_2$ planes or FeAs/FeSe layers (Fig. 1(b)).[20-22] This inherent low-dimensionality is not merely structural: it strongly constrains the electronic structure, enhances anisotropy, amplifies correlation effects, and places the pairing mechanism in an environment where competing and intertwined orders are ubiquitous.[23-27] Consequently, understanding superconductivity in the 2D limit, especially on the key superconducting planes, is central to elucidating high-$T_c$ physics. This is underscored by engineered 2D superconducting platforms such as monolayer $FeSe/SrTiO_3$[28-32] and cuprate heterostructures including $La_2CuO_4/(La,Sr)_2CuO_4$.[33,34]

In 2D superconductors, reduced screening effect enhances effective electron-electron interactions, while spontaneous symmetry breaking at interfaces or in crystals provides fertile ground for unusual superconducting phases that are difficult to realize in three-dimensional (3D) bulk materials.[35-39] These conditions prompt emergent states rooted in strong correlations, including pair density waves (PDW), a superconducting state with a spatially modulated pair amplitude that is often entangled with charge order.[40-42] In parallel, the combination of superconductivity with nontrivial band topology, frequently enabled by spin-orbit coupling and proximity effects, can yield topological superconductivity (TSC) supporting Majorana zero modes (MZMs) within magnetic vortices.[43,44] A recurring difficulty is that these phenomena are spatially localized and strongly intertwined, such that macroscopic probes can generally average out the key microscopic information.

This necessitates a tool capable of atomic-scale visualization of superconductivity. Scanning tunneling microscopy and spectroscopy (STM and STS) precisely provide this unique capability by mapping both atomic structure and the local electronic density of states (Fig. 1(c)). Its ability to correlate real-space textures with spectroscopic features has been instrumental for disentangling intertwined orders in correlated 2D superconductors.[45,46] In this review, we begin with the recent experimental advance in characterizing the superconducting planes of unconventional high-$T_c$ superconductors and tracking their doping evolution and gap structure. We then review STM/STS studies of the hot topic – PDW – and its relationship with charge order. Finally, we survey recent progress on TSC, from engineered heterostructures to intrinsic platforms, particularly on iron-based systems, highlighting both conceptual advances and material bottlenecks. We close with perspectives on major STM/STS challenges for exploring 2D superconductivity, including improving high-$T_c$ superconducting plane accessibility, underlying the microscopic mechanism of PDW and its intertwining with other orders, and progressing from spectroscopic detection to controlled manipulation of MZMs.

2. **Spectroscopic study of unconventional superconducting planes**

In cuprates and IBSCs, the essential superconducting planes ($CuO_2$ or FeAs/FeSe) are typically sandwiched between charge-reservoir or spacer layers.[5,17] This "buried-plane" architecture complicates direct measurements of intrinsic electronic properties of these planes.[7,20] Here we summarize recent experimental strategies – epitaxial growth, interface and termination control – that have overcome this limitation. We emphasize how STM/STS, with its combined atomic-scale imaging and local spectroscopy, has enabled systematic tracking of doping evolution, imposed stringent constraints on pairing symmetry, and provided direct insight into dimensional crossover toward the 2D limit in the superconducting planes.

## 2.1. CuO$_2$ plane

Cuprate high-$T_c$ superconductors share a common structural motif – the CuO$_2$ plane. In the parent compounds, the Cu$^{2+}$ ion with a 3$d^9$ electronic configuration corresponds to a half-filled band in a single-orbital picture. However, strong on-site Coulomb repulsion then drives the system into a Mott insulating state, rather than a conventional metal.[20] However, previous surface-sensitive measurements typically probe exposed charge-reservoir layers rather than the CuO$_2$ plane itself. Spectra obtained from these terminations often display rich but convoluted phenomena, including pseudogaps, competing charge orders, and debated V-shaped gap—that do not necessarily represent the intrinsic superconducting state of the buried CuO$_2$ layer layer.[21] This motivates a fundamental, direct and plane-resolved probe of the intrinsic superconducting CuO$_2$ plane. Recent advances in combining epitaxial growth and STM/STS have finally enabled such access, offering new microscopic insights of high-$T_c$ superconductivity.

An initial breakthrough was achieved using a bottom-up strategy in which a single CuO$_2$ monolayer was epitaxially grown on BiO-terminated Bi$_2$Sr$_2$CaCu$_2$O$_{8+\delta}$ (Bi-2212) surface.[47] The constructed monolayer exhibited 1 × 1 atomic lattice and a clean charge transfer gap (CTG) of approximately 2.21 eV. Unexpectedly, the subsequent STS measurements reveal a well-defined and nodeless superconducting gap of about 18.6 meV on the CuO$_2$ monolayer. This gap is robust against non-magnetic impurities (K, Cs, Ag), consistent with expectations for nodeless pairing under potential scattering as suggested by Anderson's theorem.[48] The stark contrast between V-shaped spectra on charge-reservoir surfaces (e.g., BiO) and U-shaped gaps on intrinsic CuO$_2$ terminations underscores a key interpretive principle in STS. The former, influenced by intertwined orders from charge-reservoir, may not be taken as direct proof of a nodal superconducting order within the CuO$_2$ plane. Instead, the latter offers a more faithful representation of the superconducting state, pointing towards a nodeless order parameter that merits further investigation. Our observations provided the direct spectroscopic indication that the intrinsic CuO$_2$ planes may host a fully gapped superconducting state.

To validate this observation on an intrinsic, bulk-terminated CuO$_2$ surface, attention has been turned to electron-doped infinite-layer (IL) cuprates such as SrCuO$_2$, which is naturally CuO$_2$ terminated (Fig. 2(a)).[49] On the clean CuO$_2$ surface of Sr$_{1-x}$Nd$_x$CuO$_2$ (SNCO), STS measurements also reveal a dominant U-shaped gap (Fig. 2b) with the gap closing near the superconducting transition temperature of 30 K. Beyond the coherence peaks, the spectra exhibit pronounced peak-dip-hump features, indicative of quasiparticle coupling to bosonic excitations.[50] Detailed analysis resolves three characteristic boson energies at approximately 20 meV ($\Omega_1$), 45 meV ($\Omega_2$), and 72 meV ($\Omega_3$) (Fig. 2c), which coincide with external, bending and stretching phonon modes, respectively (Fig. 2c, inset).[51] Notably, these bosonic energies remain spatially uniform despite significant variations in the local superconducting gap $\Delta$, and in some regions exceed 2$\Delta$ (Fig. 2d and 2e). Such behavior is difficult to reconcile with spin-fluctuation-mediated pairing scenarios[52] but aligns naturally with phonon-mediated coupling. These observations point to a predominantly nodeless superconducting gap on $CuO_2$ planes in both hole- and electron-doped cuprates, despite the pronounced electron-hole asymmetry of their global phase diagrams.[21]

Systematic spectroscopic study further reveals that spatial variations in local doping – quantified by shifts of the CTG midpoint ($E_F$ - $E_i$) in SNCO – from a continuous distribution (Fig. 3(a)).[49] This local doping directly modulates the superconducting gap size (Fig. 3(b)), producing a complete dome-shaped evolution within a single sample (Fig. 3(c)). Remarkably, the tunneling spectra evolve smoothly from particle-hole asymmetric gaps in underdoped regions – often associated with a Bose-Einstein-condensation limit regime – to nearly symmetric gaps near optimal doping, characteristic of more BCS-like behavior.[53] Moreover, across both regions, the superconducting gaps retain a well-defined U-shaped bottom and are accompanied by similar bosonic features.

To provide more insights in the emergence of high-$T_c$ superconductivity, systematic studies have been performed on La-doped $SrCuO_2$ (SLCO), which spans a wide doping range from hole-doped (p-SLCO) to electron-doped (n-SLCO) compositions, accompanied by an apical oxygen-induced structural transition (Fig. 4(a)).[54] The $c$ lattice constant expands from n- to p-SLCO due to the incorporation of apical oxygens at higher La concentrations (Fig. 4(b)). A key observation is that the CTG retains an approximately constant magnitude but shifts wholly relative to the Fermi level ($E_F$) (Fig. 4c), in contrast to scenarios involving strong spectral-weight transfer or Fermi-level pinning.[20,55] An even broader doping range is achieved in Eu-doped $SrCuO_2$ (SECO), owing to the smaller ionic radius of Eu.[56] Here, the superconducting transition temperature exhibits a dome-shaped dependence on Eu doping that is anti-correlated with the $c$-lattice constant. Strikingly, the CTG again remains essentially unchanged in size while shifting systematically with the $c$-axis lattice constant, closely tracking the superconducting dome. The invariance of the CTG magnitude, together with its systematic shift under both global and local doping, indicates that doping primarily modifies the electrostatic potential rather than reconstructing the underlying electronic bands. This behavior is characteristic of a modulation-doping scenario, analogous to semiconductor heterojunctions,[57] in which the $CuO_2$ plane acts as a quantum well whose carrier density is controlled by band bending induced by adjacent charge reservoir layers.[47] These findings suggest that the high-$T_c$ superconductivity in $CuO_2$ planes may arise from a strong, correlation-enhanced electron-phonon coupling mechanism embedded within a uniquely engineered 2D electronic environment, which merits further study.[58,59]

## 2.2. FeAs plane

Iron-based superconductors (IBSCs) constitute a second major family of high-$T_c$ superconductors where the essential physics resides in two-dimensional FeAs or FeSe planes.[17,18] Similar to cuprates, superconductivity in IBSCs emerges upon suppressing a correlated antiferromagnetic state –notably the nematic phase,[21,22] making them a promising platform to investigate the interplay among superconductivity, electronic correlations, nematicity and other symmetry breaking states. However, direct access to the intrinsic electronic structure of the superconducting planes, particularly the polar FeAs layers, has been hindered by surface reconstruction and disorder.[60] Therefore, the STM/STS studies predominately focused on FeSe-based systems, whether single crystals or epitaxial films.[61-64] In particular, its monolayer prepared on SrTiO$_3$ substrates exhibits interfacial superconductivity with $T_c$ exceeding 65 K and shed important light on the underlying mechanism of IBSCs.[28-32]

A decisive advance was recently achieved by surface engineering that enables STM access to the FeAs superconducting plane.[65] Specifically, a structurally matched BaAs layer was epitaxially grown on Ba(Fe$_{1-x}$Co$_x$)$_2$As$_2$ (BFCA), producing a pristine, unreconstructed surface with strongly suppressed disorder (Fig. 5(a,b)). Tunneling spectroscopy measurements on this engineered surface revealed, for the first time, spatially uniform and fully-opened U-shaped superconducting gaps in an electron-doped iron pnictide (Fig. 5(c)). The spectra are well described by a two-gap Dynes model (Fig. 5(d)), consistent with nodeless superconductivity on multiple Fermi pockets observed by ARPES.[66] Importantly, the BaAs coating layer drastically reduces quasiparticle scattering and places the system in the clean limit, enabling the observation of well-defined vortex bound states within magnetic vortex and allowing a direct microscopic determination of the superconducting order parameter. The extracted gap magnitude exhibits a dome-shaped evolution with Co doping, echoing the bulk $T_c$ phase diagram, thereby providing direct spectroscopic evidence for the doping-dependent superconducting state in the FeAs plane (Fig. 5(e)).

Spectroscopic studies on undoped FeAs surface further illuminate the correlated landscape from which superconductivity emerges. Signatures of nematicity are directly imaged as an anisotropic density of states (DOS) between neighbor atomic sites,[67] with a pronounced energy dependence and a sign reversal near 30 meV corresponding the inversion of $d_{xz}$ and $d_{yz}$ orbital splitting previously observed by ARPES.[68] In addition, a spin-density-wave (SDW) gap of approximately 25 meV is observed on pristine FeAs plane, supported by the recent observation of a similar gap on isolated FeAs region of KCa$_2$Fe$_4$As$_4$F$_2$.[69] These findings establish the FeAs plane as a clean and well-defined platform for probing the intrinsic quantum states of IBSCs and their interplay with high-$T_c$ superconductivity. Together, these results place FeSe plane on equal microscopic footing with CuO$_2$ planes and open a new path toward a unified understanding of high-$T_c$ superconductivity across different material families.

## 2.3. Fullerides

Beyond cuprates and IBSCs, alkali metal-doped fullerides ($A_x$C$_{60}$, $A$ = K, Rb, Cs) represent a distinct class of correlated superconductors with transition temperatures up to 40 K.[70] Their phase diagram exhibit hallmark features of unconventional high-$T_c$ materials, including a parent Mott insulator state, a dome-shaped superconducting region, and pseudogap behavior, yet they are largely free of the complex intertwined charge or spin orders that complicate the cuprate and IBSCs.[71] This relative simplicity, combined with intrinsically strong correlations, makes fullerides an exceptionally clean system for isolating the fundamental relationship between electron correlations and superconductivity in the 2D limit. A distinctive molecular ingredient is the dynamic

Jahn-Teller (JT) effect, which splits the orbital degeneracy and competes with Hund's coupling, thereby stabilizing a Mott-Jahn-Teller insulating ground state.[72] Despite rich physical properties, atomic-scale visualization of surface structure and superconducting gap had long remained elusive in non-cleavable bulk crystals.

A recent progress enabling microscopic insight is the successful growth of high-quality epitaxial $A_x$C$_{60}$ thin films on graphitized SiC(0001) substrate by molecular beam epitaxy (MBE).[73] This approach overcomes long-standing issues of air sensitivity and phase separation, granting STM/STS direct atomic-scale access. Crucially, it allows independent and continuous tunning of film thickness – controlling dimensionality and electronic screening – and electron doping $x$, which sets band filling and effective correlation strength.

On these epitaxial films, STM/STS measurements reveal a clear and systematic evolution of superconductivity controlled by both film thickness and electron doping. At the extreme 2D limit, monolayer and bilayer $A_3$C$_{60}$ are uniformly insulating, exhibiting large energy gaps.[73] This behavior reflects a thickness-driven SMT, where reduced screening enhances the effective $U/W$ ratio ($U$: Coulomb repulsion; $W$: band width) and stabilizes a correlated insulating state (Fig. 6(a)). Increasing thickness to trilayer and beyond restores metallicity and eventually superconductivity. Complementarily, a doping-controlled SMT is observed as well (Fig. 6(b)): increasing electron doping enhances screening, effectively reducing $U$ and driving a direct insulator-to-superconductor transition even in bilayer K$_x$C$_{60}$ films, closely resembling the doping-driven phase diagrams of cuprates and IBSCs.

Spectroscopically, the superconducting state is characteristic of an isotropic $s$-wave order parameter.[73] This conclusion is confirmed by consistently observing U-shaped tunneling spectra (Fig. 6(c,d)), the absence of in-gap states at nonmagnetic impurities and step edges (Fig. 6(e)),[74] and the appearance of Yu-Shiba-Rusinov (YSR) states only around magnetic adatoms. Notably, superconductivity was also found to coexist with a pseudogap-like phase, a phenomenon reminiscent of cuprates. The pairing is found to be remarkably robust in K$_3$C$_{60}$ and Rb$_3$C$_{60}$, even though these materials reside in a strong correlation regime ($U/W \approx 3$). It suggests that in this narrow-band system, strong electronic correlations do not destroy the $s$-wave order but may instead cooperate with local Jahn-Teller phonons to stabilize a local, correlation-enhanced pairing.

Combining thickness- and doping-dependent measurements yields a unified phase diagram: both the superconducting gap $\Delta$ and $T_c$ exhibit a dome-shaped dependence on electron doping, peaking precisely at half-filling ($x = 3$).[73] This behavior cannot be explained by a simple density-of-states-driven BCS mechanism. Instead, it points to a local pairing mechanism mediated by high-frequency intramolecular JT phonons, in which strong correlations and JT physics cooperate to form spin-singlet pairs on individual C$_{60}$ molecules. The extremely short coherence length of a 1.5~2.6 nm— comparable to only two intermolecular spacings—provides direct evidence for real-space, short-range pairing. The superconducting dome naturally arises because electronic correlations initially enhance pairing but eventually localize carriers and suppress superconductivity when they become too strong, as in the MJTI phase. As such, the fulleride platform provides a valuable reference system with broad implications for understanding pairing mechanism in other narrow-band, strongly correlated superconductors.[75]

### 3. Pair-density wave

The PDW represents an exotic superconducting state in which the Cooper-pair condensate acquires a finite wavevector, leading to a spatial modulation of the superconducting order parameter which is fundamentally distinct from a uniform BCS superconductivity.[40] While the original theoretical proposal – the Fulde-Ferrell-Larkin-Ovchinnikov (FFLO) state – relies on Zeeman-split Fermi surface and typically appears under high magnetic fields,[76] growing experimental evidence has established a different class of PDWs driven primarily by electronic correlations at zero field.[77] Such correlation-driven PDW has now been identified across multiple families of unconventional superconductors, including cuprates, Kagome metals, IBSCs, and transition-metal dichalcogenides (TMDCs). A crucial feature of these discoveries, enabled largely by STM/STS and scanned Josephson tunneling microscopy (SJTM), is the intimate and often non-trivial entanglement between PDW and pre-existing charge density wave (CDW) order. Understanding this relationship is central to unraveling intertwined orders and the enigmatic pseudogap phase in unconventional high-$T_c$ superconductors.

A key physical question underpinning the phenomenology is the relationship between the observed PDW and pre-existing CDW orders: which one is the primary driver? Theoretically, the two orders can be intimately coupled. A primary CDW may induce a PDW by coupling to a uniform superconducting component. Conversely, a primary PDW state can induce charge modulations at a wavevector twice of its characteristic wavevector. STM/S plays a decisive role in constraining these models by providing real-space information on the wavevectors, spatial coherence, and stability of both modulations. As discussed below, the detection of secondary CDW at specific harmonics of the PDW wavevector provides strong evidence for a primary PDW scenario in unconventional superconductors.

In cuprates, PDW physics is of particular interest due to its proposed connection to the intensely-discussed pseudogap in underdoped regime.[77] Direct phase-sensitive evidence for PDW correlations was firstly obtained by scanned Josephson tunnelling microscopy (SJTM) on optimally doped $Bi_2Sr_2CaCu_2O_{8+\delta}$ (Bi2212) using a superconducting Bi2212 nanoflake tip.[78] In these measurements, the magnitude of the Josephson critical current $|I_J(r)|$ exhibits a spatial modulation with a period of $4a_0$. As clarified in later analyses, owing to the extended nature of the nanoflake tip and momentum-conserving planar tunnelling, an intrinsic PDW with an $8a_0$ modulation of the pairing amplitude naturally gives rise to a $4a_0$ modulation in $|I_J(r)|$.[79,80] Consistently, a PDW order at $Q_P \sim 2\pi/8a_0$ and its secondary charge modulation at $2Q_P \sim 2\pi/4a_0$ are simultaneously observed at vortex halo under magnetic field, as expected when a primary PDW coexists with a uniform superconducting condensate.[81] These results establish that the PDW as an intrinsic pairing instability of Bi2212, coexisting with a uniform superconducting order parameter. Under applied magnetic fields, the same PDW is strongly enhanced in vortex halos, where uniform superconductivity is locally suppressed.

Beyond SJTM, single-particle STM spectroscopy provides complementary evidence for PDW order.[82] In underdoped Bi-2212, universal real-space modulations of the coherence-peak height (Fig. 7(a)) and gap depth (Fig. 7(b)) with a period of $4a_0$ were observed, directly correlating with the coupling of the $4a_0$ plaquette charge order and uniform superconductivity (Fig. 7(c)). Upon overdoping, both the $4a_0$ CDW and the associated PDW disappear simultaneously (Fig. 7(d)), accompanied by the emergence of a short-range $\sqrt{2}\times\sqrt{2}$ order (Fig. 7(e)).[82] The systematic doping evolution — where the $4a_0$ CDW and its associated PDW disappear in the overdoped regime concurrently with the pseudogap, provides empirical and phenomenological support for their intertwined role in pseudogap physics.[42]

Materials with robust CDW order provide a fertile ground for exploring PDW and its interplay with intertwined orders beyond cuprates. For example, in Kagome superconductor $CsV_3Sb_5$, SJTM measurements revealed a primary 3Q PDW with a $4a_0/3 \times 4a_0/3$ periodicity, coexisting with unidirectional $4a_0$ charge stripes and a $2a_0 \times 2a_0$ charge order (Fig. 8(a-d)).[83] This novel PDW order and its secondary 3Q charge order persists under strong magnetic field and above $T_c$ within the energy scale of ±5 meV, pointing to a primary pairing modulation associated with a pseudogap-like feature, closely analogous to underdoped cuprates.

More remarkably, the PDW in $KV_3Sb_5$ displays field-tunable chirality with a $2a_0 \times 2a_0$ periodicity (Fig. 8(e)).[84] Theoretically, this behavior is attributed to interband pairing between Sb-$p$ and V-$d$ orbitals that leads to chiral finite-momentum Cooper pairs (Fig. 8(f,g)). Chemical substitution further enriches the PDW physics: in Ta-doped $CsV_3Sb_5$, suppression of the pristine $2a_0 \times 2a_0$ CDW coincides with enhanced $T_c$ and the emergence of a time-reversal symmetry breaking (TRSB) superconducting state, as evidenced by muon spin resonance ($u$SR) measurements.[85] Combined STM and quasiparticle interference (QPI) measurements reveal gap modulations at a incommensurate ring-like wavevector $Q_a$ linked to interband scattering, underscoring the sensitivity of PDW order to band topology and chemical doping, which warrants further investigation.

Atomic-scale surface engineering offers an additional control knob.[86] The deliberate fabrication of a Cs 2×2 overlayer on Kagome Sb surface stabilizes a novel 2D superconducting state with a significantly enhanced $T_c$ and emergent surface bands. The emergence and on/off switching of this superconducting state depend critically on the dimension and registry (in-phase or out-of-phase stacking) of the 2×2 Cs layer. Notably, a robust $2a_0 \times 2a_0$ PDW is established within this engineered phase, providing a clean and tunable platform to dissect the CDW-PDW-superconductivity interplay.

As prototypical 2D correlated materials, TMDCs provide another complementary perspective. In the superconducting state of $NbSe_2$, a $3a_0 \times 3a_0$ PDW emerges at the same wavevectors as the pre-existing CDW, but with a nontrivial phase shift between them of $\sim 2\pi/3$.[87] Detailed STM analysis shows that different CDW registries (anion-centered versus hollow-centered) generate distinct PDW configurations, with anion-centered regions showing spontaneous rotational symmetry breaking—a signature of a smectic PDW state.[88] Similar lock-in relationship between PDW and CDW has also been observed in heavy fermion superconductor UTe$_2$, where the superconducting gap modulates with CDW period and exhibits a relative phase shift of π.[89] Meanwhile, the unidirectional PDW orders have been increasingly evidenced in EuRbFe$_4$As$_4$, FeTe$_{0.5}$Se$_{0.5}$, and slightly electron-doped topological monolayer 1T′-MoTe$_2$.[90-92] Nevertheless, these PDW modulations observed exhibit a periodicity several times larger than the atomic lattice constant.

At the ultimate limit of spatial resolution, recent STM/STS studies have uncovered PDW modulations within a single crystallographic unit cell and even s sublattice length. Upon moderate doping of monolayer 1T′-MoTe$_2$, the emergence of an additional electron pocket along the ΓX direction induces completely new PDW and CDW orders along the *a*-axis (Fig. 9(a)).[93] The superconducting gap exhibits a maximum on the underlying Te sublattice and oscillates with a half-unit-cell periodicity (Fig. 9(b,c)). Comparable intra-unit-cell PDW modulations causes the gap maximum to alternate between Te/Se sites.[94] In thicker Fe(Te, Se) flakes, the superconducting gap maximum shifts to Fe sites, producing an intra-unit-cell PDW modulation as well.[95]

Across these material families, the investigation of PDW was often motivated by its theoretical connection to the cuprate pseudogap. The link is somewhat supported by their experimental phase coexistence and the persistence of PDW-like states above $T_c$ in Kagome superconductor AV$_3$Sb$_5$, an intriguing phenomenological parallel with cuprates that warrants further study to establish universal microscopic picture of pseudogap. Nevertheless, these observations demonstrate that coupling between superconductivity, lattice symmetry, and orbital degrees of freedom can reshape the superconducting order parameter at the most microscopic scale of crystal lattice.

## 4. Topological superconductivity

Topological superconductivity, a distinct class of exotic phase, emerges when superconducting pairing develops in a system with nontrivial band topology. This phase is of central interest as it can host localized MZMs bound to topological defects such as vortices, edges, or atomic-scale defects, providing a promising route toward fault-tolerant topological quantum computation. Seminal proposals, exemplified by the Fu-Kane model,[96] showed that proximity-coupling between a conventional s-wave superconductor and a topological insulator can induce an effective topological superconducting state. In this context, STM/STS has played an outsized role by providing local, energy-resolved fingerprints– most notably zero-bias-conductance peaks (ZBCPs) and their spatial variation–needed to identify candidate MZMs in magnetic vortices. The following sections survey key STM/STS advances in visualizing 2D topological superconductivity and its harboring MZMs, from heterostructure-based realizations to intrinsic material platforms unifying superconductivity and nontrivial band topology in a single compound.

### 4.1. Hybrid heterostructures

Early efforts toward TSC firstly focused on heterostructures in which an *s*-wave superconductor supplies the pairing potential to a low-dimensional system with strong SOC, such as growth of three-dimensional topological insulator (TI) on NbSe$_2$. This realizes the well-known Fu-Kane-type physics. STM/STS on $Bi_2Te_3$/$NbSe_2$ has observed long, non-splitting ZBCPs within vortices for TI films thicker than three quintuple layers, consistent with MZM expectations (Fig. 10(a,b)).[97] The approach has been extended to topological crystalline insulators such as SnTe, where crystalline symmetries may allow multiple MZMs per vortex.[98] In SnTe/$NbSe_2$, vortex-core-zero-bias features can extend over unusually long distances (~ 100 nm) when the magnetic field aligns with a (110)-type mirror plane, otherwise splitting due to MZM hybridization (Fig. 10(c-e)). Atomically thin Bi films, proposed as 2D topological insulators, have also been successfully proximitized by $NbSe_2$,[99] providing a route to combining topological edge states with superconductivity under time-reversal-symmetry breaking fields.

A major limitation of $NbSe_2$-based heterostructures is their low operating temperature ($T_c$ < 7 K). To push toward higher temperatures, attention has been paid on heterostructures employing high-$T_c$ superconductors as the pairing source. A representative example is the MnTe/$Bi_2Te_3$/Fe(Te, Se) trilayer, integrating magnetism, TI surface states, and an iron-based superconductor ($T_c$ ~ 14 K).[100] STM/STS on the topmost MnTe layer reveals a sizable proximity-induced U-shaped gap (~3 meV) and a high density of nanoscale regions (~ 4-10 nm), showing robust, field-insensitive ZBCPs, suggestive of possible MZMs. Closely related concepts include interfacing the topological magnet $MnBi_2Te_4$ with Fe(Te, Se), where the interplay between magnetism and superconductivity can be tuned via layer thickness.[101] Pushing further, ultrathin bismuth films on the cuprate Bi-2212 ($T_c$ > 90 K) exhibit proximity gaps as large as 7.5 meV, opening an appealing but still experimentally challenging route to high-$T_c$ TSC, for which definitive MZM evidence remains a future challenge.[102]

**4.2. Connate topological superconductors**

While heterostructures established proof-of-principle routes to TSC, their progress is often constrained by interface disorder, material complexity, and limited operating temperatures. In parallel, increasing attention has also been paid to search for connate topological superconductors hosting coexisting superconductivity and nontrivial band topology, such as $Cu_xBi_2Se_3$, β-$Bi_2Pd$ and 2M-$WS_2$.[103-105] In this context, a transformative advance came with the observation of TSC in IBSCs,[106], arising from band inversion involving $p$ orbitals of the chalcogen/pnictogen layers and $d$ orbitals of the iron layers (Fig. 11(a)).[107] This intrinsic integration eliminates the need for delicate proximity engineering and has positioned IBSCs, particularly Fe(Te, Se), at the forefront of TSC research.

The initial breakthrough was made in $FeTe_{0.55}Se_{0.45}$, where Te/Se substitution tunes the electronic structure to host topological surface states, as established by ARPES. STM/S measurements revealed robust ZBCP localized at vortex cores, consistent with vortex-bound MZMs (Fig. 11(b,c)).[108] Subsequent experiments strengthened this interpretation, including the observation of a quantized tunneling conductance plateau at $2e^2/h$, which provides a compelling signature of evidence of a particle-hole-symmetric zero mode. In the related compound $(Li_{0.84}Fe_{0.16})OHFeSe$ ($T_c$ ~ 42 K), the larger superconducting gap and improved quantum-limit conditions yield clearer spectra in which the MZM is well separated from higher-energy Caroli-de Gennes-Matricon (CdGM) states (Fig. 11(d)).[109] A conceptual expansion followed with the realization of possible MZMs bound to topological defects even without an external magnetic field. In monolayer FeTe$_{0.5}$Se$_{0.5}$ film, ZBCPs have been observed at the ends of one-dimensional atomic line defects and survives above 20 K, remaining robust under applied magnetic fields.[110] Furthermore, depositing individual magnetic Fe adatoms on Fe(Te, Se) can nucleate localized quantum anomalous vortices that trap MZMs, suggesting an STM-enable route toward atomic-scale creation of manipulation of topological defects.[111]

The most recent landmark is the realization of an ordered and tunable MZM lattice in LiFeAs. Early STM studies did not observe such MZMs on pristine LiFeAs, contrary to the theoretically predicted topological bands there. The progress was found in the need to tune chemical potential and reduce rotational symmetry.[112] In systems hosting a naturally occurring strain-induced biaxial CDW, the CDW simultaneously provides a periodic vortex-pinning potential and gaps out the topological Dirac semimetal state, stabilizing a topological superconducting phase.[113] Under magnetic field, vortices self-assemble into an ordered lattice pinned along CDW stripes, with over 90% of vortices hosting clear MZM signatures (Fig. 11(e)). This scalable MZM array marks a shift from sporadic detection toward partial controllability and material-enabled design.

Despite this rapid progress, material challenges remain central. In Fe(Te, Se), intrinsic disorder can strongly limit the yield of vortices hosting MZMs. For the other IBSCs such as prominent 122 iron-pnictides (e.g., BaFe$_2$As$_2$), predicted topological surface states remain difficult to access by STM due to surface reconstruction and disorder upon cleavage. To date, CaKFe$_4$As$_4$ represents a notable case in which alternative stacking of Ca and K layers mitigates these issues and enable MZM observation.[114]

## 5. Conclusion and perspectives

In this review, we discuss about how STM/STS has advanced our understanding of 2D superconductivity from phenomenology to microscopic, plane-resolved investigation. By directly accessing superconducting CuO$_2$ and FeAs layers that are usually buried beneath reservoir or spacer blocks, STM/STS has connected crystal structure, local electronic spectra, and macroscopic superconducting behavior with atomic precision. Several common themes emerge across different material families. First, direct STM/STS measurements frequently reveal fully gapped spectra on superconducting planes of cuprates, IBSCs, and molecular fullerides. Together with plane-resolved doping evolution and ubiquitous electron-boson features in multiple systems, these observations motivate renewed scrutiny of how low dimensionality, electrostatics, and correlations jointly shape pairing.[115] A key challenge is to reconcile plane-resolved spectroscopic constraints with phase-sensitive evidence and momentum-space information, and to clarify under what conditions nodeless spectra imply specific pairing symmetries.

In addition, PDW physics appears as recurring instability in correlated 2D

superconductors. Across cuprates, Kagome metals and IBSCs, PDW signatures are typically intertwined with charge order, nematicity, and lattice-scale symmetry breaking, sometimes extending down to intra-unit-cell modulations. Establishing a unified framework that links PDW, CDW, and pseudogap phenomena – while distinguishing primary from induced orders – remains an open and timely direction. Meanwhile, TSC research has gained progress in both proximity-engineered heterostructures to intrinsic systems, with IBSCs offering a rare combination of high-$T_c$ superconductivity and nontrivial band topology. STM/STS visualization of MZMs in vortices, line defects, and adatoms, together with progress toward Majorana arrays, marks a transition from sporadic detection toward partial controllability. The next step is to improve materials homogeneity and reproducibility, and to develop experimentally practical protocols that move beyond identification to manipulation.

Looking forward, progress will increasingly depend on the ability to prepare ideal superconducting planes and interfaces with atomic precision. Interface and epitaxial design – exemplified by BaAs/BFCA - provides a powerful route to overcome surface disorder and reconstruction, enabling systematic studies of intertwined and topological states in cleaner environment. Ultimately, continued synergy among atomically controlled synthesis, advanced local probes, and theory will be essential to transform 2D superconductivity from discovery-driven exploration into design-driven quantum matter engineering.[116,117]

**Acknowledgment**

This work was financially supported by the National Natural Science Foundation of China (Grant Nos. 12474130, 12141403, 12134008) and the National Key R&D Program of China (Grant Nos. 2022YFA1403100).

**References**

[1] Onnes H K 1911 *Comm. Phys. Lab. Univ. Leiden* **12** 120

[2] Meissner W and Ochsenfeld R 1933 *Naturwiss.* **21** 787

[3] Yao C and Ma Y W 2021 *iScience* **24** 102541

[4] Bardeen J, Cooper L N, and Schrieffer J R 1957 *Phys. Rev.* **108** 1175

[5] Bednorz J G and Müller K A 1986 *Z. Phys. B: Condens. Matter* **64** 189

[6] Timusk T and Statt B W 1999 *Rep. Prog. Phys.* **62** 61

[7] Damascelli A, Hussain Z, and Shen Z X 2003 *Rev. Mod. Phys.* **75** 473

[8] Fong H F, Keimer B, Anderson P W, Reznik D, Dogan F, and Aksay I A 1995 *Phys. Rev. Lett.* **75** 316

[9] Loeser A G, Shen Z X, Dessau D S, Marshall D S, Park C H, Fournier P, and Kapitulnik A 1996 *Science* **273** 325

[10] Valla T, Fedorov A V, Johnson P D, Glans P A, McGuinness C, Smith K E, Andrei E Y, and Berger H 2004 *Phys. Rev. Lett.* **92** 086401

[11] Xu Z A, Ong N P, Wang Y, Kakeshita T, and Uchida S 2000 *Nature* **406** 486

[12] Lanzara A, Bogdanov P V, Zhou X J, Kellar S A, Feng D L, Lu E D, Yoshida T, Eisaki H, Fujimori A, Kishio K, Shimoyama J I, Noda T, Uchida S, Hussain Z, and Shen Z X 2001 *Nature* **412** 510

[13] Pan S H, O'Neal J P, Badzey R L, Chamon C, Ding H, Engelbrecht J R, Wang Z, Eisaki H,

Uchida S, Gupta A K, Ng K W, Hudson E W, Lang K M, and Davis J C 2001 *Nature* **413** 282
[14] Hoffman J E, McElroy K, Lee D H, Lang K M, Eisaki H, Uchida S, and Davis J C 2002 *Science* **297** 1148
[15] Anderson P W 1987 *Science* **235** 1196
[16] Sheng D N, Chen Y C, and Weng Z Y 1996 *Phys. Rev. Lett.* **77** 5102
[17] Kamihara Y, Watanabe T, Hirano M, and Hosono H 2008 *J. Am. Chem. Soc.* **130** 3296
[18] Dai P, Hu J, and Dagotto E 2012 *Nat. Phys.* **8** 709
[19] Hosono H, Yamamoto A, Hiramatsu H, and Ma Y 2018 *Mater. Today* **21** 278
[20] Lee P A, Nagaosa N, and Wen X G 2006 *Rev. Mod. Phys.* **78** 17
[21] Keimer B, Kivelson S A, Norman M R, Uchida S, and Zaanen J 2015 *Nature* **518** 179
[22] Paglione J and Greene R L 2010 *Nat. Phys.* **6** 645
[23] Scalapino D J 2012 *Rev. Mod. Phys.* **84** 1383
[24] Wen X G and Lee P A 1996 *Phys. Rev. Lett.* **76** 503
[25] Dagotto E 1994 *Rev. Mod. Phys.* **66** 763
[26] Shen Z X, Dessau D S, Wells B O, Olson C G, Mitzi D, Lindau I, Spicer W E, and Kapitulnik A 1993 *Phys. Rev. Lett.* **70** 1553
[27] Stewart G R 2011 *Rev. Mod. Phys.* **83** 1589
[28] Wang Q Y, Li Z, Zhang W H, Zhang Z C, Zhang J S, Li W, Ding H, Ou H W, Deng P, Chang K, Wen J, Song C L, He K, Jia J F, Ji S H, Wang Y Y, Wang L L, Chen X, Ma X C, and Xue Q K 2012 *Chin. Phys. Lett.* **29** 037402 (in Chinese)
[29] He S L, He J F, Zhang W H, et al. 2013 *Nat. Mater.* **12** 605
[30] Ge J F, Liu Z L, Liu C, Gao C L, Qian D, Xue Q K, Liu Y, and Jia J F 2015 *Nat. Mater.* **14** 285
[31] Lee J J, Schmitt F T, Moore R G, Johnston S, Cui Y T, Li W, Yi M, Liu Z K, Hashimoto M, Zhang Y, Lu D H, Devereaux T P, Lee D H, and Shen Z X 2014 *Nature* **515** 245
[32] Xu Y, Rong H, Wang Q, Ma Z, Cai P, Ge J F, Wang C, Zhang Y, Gu L, Liu K, Wen H H, and Xue Q K 2021 *Nat. Commun.* **12** 2840
[33] Gozar A, Logvenov G, Kourkoutis L F, Bollinger A T, Giannuzzi L A, Muller D A, and Bozovic I 2008 *Nature* **455** 782
[34] Shen J Y, Shi C Y, Pan Z M, *et al.* 2023 *Nat. Commun.* **14** 7290
[35] Jungfleisch M B, Zhang W, Sklenar J, Jiang W, Pearson J E, Ketterson J B, and Hoffmann A 2016 *Phys. Rev. B* **93** 224419
[36] Presland M R, Tallon J L, Buckley R G, Liu R S, and Flower N E 1991 *Physica C* **176** 95
[37] Emery V J and Kivelson S A 1995 *Nature* **374** 434
[38] Novoselov K S, Mishchenko A, Carvalho A, and Castro Neto A H 2016 *Science* **353** aac9439
[39] Schaibley J R, Yu H, Clark G, Rivera P, Ross J S, Seyler K L, Yao W, and Xu X 2016 *Nat. Rev. Mater.* **1** 16055
[40] Agterberg D F, Davis J S, Edkins S D, et al. 2020 *Annu. Rev. Condens. Matter Phys.* **11** 231
[41] Fradkin E, Kivelson S A, and Tranquada J M 2015 *Rev. Mod. Phys.* **87** 457
[42] Lee P A 2014 *Phys. Rev. X* **4** 031017
[43] Kitaev A Yu 2001 *Phys.-Usp.* **44** 131
[44] Kitaev A Y 2003 *Ann. Phys.* **303** 2
[45] Fischer Ø, Kugler M, Maggio-Aprile I, Berthod C, and Renner C 2007 *Rev. Mod. Phys.* **79** 353
[46] Yazdani A, da Silva Neto E H, and Aynajian P 2016 *Annu. Rev. Condens. Matter Phys.* **7** 11
[47] Zhong Y, Wang Y, Han S, Lv Y F, Wang W L, Zhang D, Ding H, Zhang Y M, Wang L L, He

K, Zhong R D, Schneeloch J A, Gu G D, Song C L, Ma X C, and Xue Q K 2016 *Sci. Bull.* **61** 1239 (in Chinese)

[48] Anderson P W 1959 *J. Phys. Chem. Solids* **11** 26

[49] Fan J Q, Yu X Q, Cheng F J, Wang H, Wang R F, Ma X B, Hu X P, Zhang D, Ma X C, Xue Q K, and Song C L 2022 *Natl. Sci. Rev.* **9** nwab225 (in Chinese)

[50] Niestemski F C, Kunwar S, Zhou S, et al. 2007 *Nature* **450** 1058

[51] Tajima S, Ido T, Ishibashi S, Itoh T, Eisaki H, Mizuo Y, Arima T, Takagi H, and Uchida S 1991 *Phys. Rev. B* **43** 10496

[52] Eschrig M 2006 *Adv. Phys.* **55** 47

[53] Zhu Q, Fan J Q, Yu X Q, Xiong Y L, Yan H, Wang R F, Song C L, Ma X C, and Xue Q K 2025 *Phys. Rev. B* **111** 245410

[54] Zhong Y, Fan J Q, Wang R F, Wang S Z, Zhang X F, Zhu Y Y, Dou Z Y, Yu X Q, Wang Y, Zhang D, Zhu J, Song C L, Ma X C, and Xue Q K 2020 *Phys. Rev. Lett.* **125** 077002

[55] Molegraaf H J A, Presura C, Van Der Marel D, Kes P H, and Li M 2002 *Science* **295** 2239

[56] Yu X Q, Yan H, Wei L X, Deng Z X, Xiong Y L, Fan J Q, Yu P, Ma X C, Xue Q K, and Song C L 2022 *Phys. Rev. B* **106** L100503

[57] Dingle R, Störmer H L, Gossard A C, and Wiegmann W 1978 *Appl. Phys. Lett.* **33** 665

[58] Song C L, Ma X C, and Xue Q K 2020 *MRS Bull.* **45** 366

[59] Wang R F, Song C L, Ma X C, and Xue Q K 2025 *AAPPS Bull.* **35** 12

[60] Yin Y, Zech M, Williams T L, Wang X F, Wu G, Chen X H, and Hoffman J E 2009 *Phys. Rev. Lett.* **102** 097002

[61] Wei Z, Qin S, Ding C, Wu X, Hu J, Sun Y J, Zhao L X, Li J, Zhang Y, He K, Xue Q K, and Chen X 2023 *Nat. Commun.* **14** 5302

[62] Yuan Y, Fan X, Wang X, He K, Zhang Y, Xue Q K, and Li W 2021 *Nat. Commun.* **12** 2196

[63] Song C L, Wang Y L, Cheng P, et al. 2011 *Science* **332** 1410

[64] Song C L, Zhang H M, Zhong Y, Hu X P, Ji S H, Wang L L, He K, Ma X C, and Xue Q K 2016 *Phys. Rev. Lett.* **116** 157001

[65] Ren M Q, Cheng Q J, He H H, Deng Z X, Cheng F J, Wang Y W, Lou C C, Zhang Q H, Gu L, Liu K, Ma X C, Xue Q K, and Song C L 2025 *Phys. Rev. Lett.* **134** 246203

[66] Terashima K, Sekiba Y, Bowen J H, Nakayama K, Kawahara T, Sato T, Richard P, Uchiyama H, Souma S, Sato T, Takahashi T, and Komatsu H 2009 *Proc. Natl. Acad. Sci. USA* **106** 7330

[67] Cheng Q J, Wang Y W, Ren M Q, Deng Z X, Lou C C, Ma X C, Xue Q K, and Song C L 2025 *Commun. Mater.* **6** 162

[68] Yi M, Lu D, Chu J H, Analytis J G, Sorini A P, Kemper A F, Moritz B, Mo S K, Moore R G, Hashimoto M, Lee W S, Hussain Z, Devereaux T P, Fisher I R, and Shen Z X 2011 *Proc. Natl. Acad. Sci. USA* **108** 6878

[69] Shao S, Zhang F, Zhang Z Y, Wang T, Wu Y W, Tu Y B, Hou J, Hou X Y, Hao N, Mu G, and Shan L 2023 *Sci. China-Phys. Mech. Astron.* **66** 287412 (in Chinese)

[70] Fleming R M, Ramirez A P, Rosseinsky M J, Murphy D W, Haddon R C, Zahurak S M, and Makhija A V 1991 *Nature* **352** 787

[71] Takabayashi Y and Prassides K 2016 *Phil. Trans. R. Soc. A* **374** 20150320

[72] Han S, Guan M X, Song C L, Wang Y L, Ren M Q, Meng S, Ma X C, and Xue Q K 2020 *Phys. Rev. B* **101** 085413

[73] Ren M Q, Han S, Wang S Z, Fan J Q, Song C L, Ma X C, and Xue Q K 2020 *Phys. Rev.*

*Lett.* **124** 187001

[74] Wang S Z, Ren M Q, Han S, Cheng F J, Ma X C, Xue Q K, and Song C L 2021 *Commun. Phys.* **4** 114

[75] Ren M Q, Wang S Z, Han S, Song C L, Ma X C, and Xue Q K 2022 *AAPPS Bull.* **32** 1

[76] Fulde P and Ferrell R A 1964 *Phys. Rev.* **135** A550

[77] Chen H D, Vafek O, Yazdani A, and Zhang S C 2004 *Phys. Rev. Lett.* **93** 187002

[78] Hoffman J E, Hudson E W, Lang K M, Madhavan V, Eisaki H, Uchida S I, and Davis J C 2002 *Science* **295** 466

[79] Edkins S D, Kostin A, Fujita K, Mackenzie A P, Eisaki H, Uchida S, Sachdev S, Lawler M J, Kim E A, Davis J C S, and Hamidian M H 2019 *Science* **364** 976

[80] Du Z, Li H, Joo S H, Donoway E P, Lee J, Davis J C S, Gu G, Johnson P D, and Fujita K 2020 *Nature* **580** 65

[81] Li X, Zou C, Ding Y, Yan H, Ye S, Li H, Hao Z Q, Zhao L, Zhou X J, and Wang Y Y 2021 *Phys. Rev. X* **11** 011007

[82] Ruan W, Li X, Hu C, Hao Z Q, Li H W, Cai P, Zhou X J, Lee D H, and Wang Y Y 2018 *Nat. Phys.* **14** 1178

[83] Chen H, Yang H T, Hu B, et al. 2021 *Nature* **599** 222

[84] Deng H, Qin H, Liu G, et al. 2024 *Nature* **632** 775

[85] Deng H, Liu G, Guguchia Z, et al. 2024 *Nat. Mater.* **23** 1639

[86] Han X, Chen H, Tan H, Cao Z, Huang Z, Ye Y, Zhao Z, Shen C, Yang H, Yan B, Wang Z, and Gao H J 2025 *Nat. Nanotechnol.* **20** 1017

[87] Liu X, Chong Y X, Sharma R, and Davis J S 2021 *Science* **372** 1447

[88] Cao L, Xue Y, Wang Y, Zhang F C, Kang J, Gao H J, Mao J, and Jiang Y 2024 *Nat. Commun.* **15** 7234

[89] Gu Q, Carroll J P, Wang S, Ran S, Broyles C, Siddiquee H, Butch N P, Saha S R, Paglione J, Davis J C S, and Liu X 2023 *Nature* **618** 921

[90] Zhao H, Blackwell R, Thinel M, Handa T, Ishida S, Zhu X, Iyo A, Eisaki H, Pasupathy A N, and Fujita K 2023 *Nature* **618** 940

[91] Liu Y, Wei T, He G, Zhang Y, Wang Z, and Wang J 2023 *Nature* **618** 934

[92] Wei L X, Xiao P C, Li F S, Wang L, Deng B Y, Cheng F J, Zheng F W, Hao N, Zhang P, Ma X C, Xue Q K, and Song C L 2025 *Phys. Rev. B* **112** L060503

[93] Cheng F J, Lou C C, Chen A X, Wei L X, Liu Y, Deng B Y, Li F, Wang Z, Xue Q K, Ma X C, and Song C L 2025 *Phys. Rev. Lett.* **135** 166201

[94] Wei T, Liu Y, Ren W, Liang Z, Wang Z, and Wang J 2025 *Chin. Phys. Lett.* **42** 027404 (in Chinese)

[95] Kong L, Papaj M, Kim H, Zhang Y, Baum E, Li H, Watanabe K, Taniguchi T, Gu G, Lee P A, and Nadj-Perge S 2025 *Nature* **640** 55

[96] Fu L and Kane C L 2008 *Phys. Rev. Lett.* **100** 096407

[97] Kezilebieke S, Huda M N, Vaňo V, Aapro M, Ganguli S C, Silveira O J, Głodzik S, Foster A S, Ojanen T, and Liljeroth P 2020 *Nature* **588** 424

[98] Xu J P, Wang M X, Liu Z L, Ge J F, Yang X, Liu C, Xu Z A, Guan D, Gao C L, Qian D, Liu Y, Wang Q H, Zhang F C, Xue Q K, and Jia J F 2015 *Phys. Rev. Lett.* **114** 017001

[99] Liu T, Wan C Y, Yang H, Zhao Y J, Xie B J, Zheng W Y, Yi Z X, Guan D D, Wang S Y, Zheng H, Liu C H, Fu L, Liu J W, Li Y Y, and Jia J F 2024 *Nature* **633** 71

[100]  Sun H H, Wang M X, Zhu F, Wang G Y, Ma H Y, Xu Z A, Liao Q, Lu Y, Gao C L, Li Y Y, Liu C, Qian D, Guan D, and Jia J F 2017 *Nano Lett.* **17** 3035

[101]  Ding S, Chen C, Cao Z, Wang D, Pan Y, Tao R, Zhao D, Hu Y, Jiang T, Yan Y, Shi Z, Wan X, Feng D, and Zhang T 2022 *Sci. Adv.* **8** eabq4578

[102]  Yuan W, Yan Z J, Yi H, Wang Z, Paolini S, Zhao Y F, Zhou L, Wang A G, Wang K, Prokscha T, Salman Z, Suter A, Balakrishnan P P, Grutter A J, Winter L E, Singleton J, Chan M H W, and Chang C Z 2024 *Nano Lett.* **24** 7962

[103]  Yuan Y H, Pan J, Wang X T, Fang Y Q, Song C L, Wang L L, He K, Ma X C, Zhang H J, Huang F Q, Li W, and Xue Q K 2019 *Nat. Phys.* **15** 1046

[104]  Fan X M, Sun X Q, Zhu P H, Fang Y Q, Ju Y K, Yuan Y H, Yan J M, Huang F Q, Hughes T L, Tang P Z, Xue Q K, and Li W 2025 *Natl. Sci. Rev.* **12** nwae312 (in Chinese)

[105]  Lv Y F, Wang W L, Zhang Y M, Ding H, Li W, Wang L L, He K, Song C L, Ma X C, and Xue Q K 2017 *Sci. Bull.* **62** 852 (in Chinese)

[106]  Wu X, Liu X, Thomale R, and Liu C X 2022 *Natl. Sci. Rev.* **9** nwab087 (in Chinese)

[107]  Zhang P, Yaji K, Hashimoto T, Ota Y, Kondo T, Okazaki K, Wang Z, Wen J, Gu G D, Ding H, and Shin S 2018 *Science* **360** 182

[108]  Wang D, Kong L, Fan P, Chen H, Zhu S, Liu W, Cao L, Sun Y, Du S, Schneeloch J, Zhong R, Gu G, Fu L, Ding H, and Gao H J 2018 *Science* **362** 333

[109]  Liu Q, Chen C, Zhang T, Peng R, Yan Y J, Wen C H P, Lou X, Huang Y L, Tian J P, Dong X L, Wang G W, Bao W C, Wang Q H, Yin Z P, Zhao Z X, and Feng D L 2018 *Phys. Rev. X* **8** 041056

[110]  Chen C, Jiang K, Zhang Y, Liu C, Liu Y, Wang Z, and Wang J 2020 *Nat. Phys.* **16** 536

[111]  Fan P, Yang F, Qian G, Chen H, Zhang Y Y, Li G, Huang Z, Xing Y, Kong L, Liu W, Jiang K, Shen C, Du S, Schneeloch J, Zhong R, Gu G, Wang Z, Ding H, and Gao H J 2021 *Nat. Commun.* **12** 1348

[112]  Kong L, Cao L, Zhu S, Papaj M, Dai G, Li G, Fan P, Liu W, Yang F, Wang X, Du S, Jin C, Fu L, Gao H J, and Ding H 2021 *Nat. Commun.* **12** 4146

[113]  Li M, Li G, Cao L, Zhou X, Wang X, Jin C, Chiu C K, Pennycook S J, Wang Z, and Gao H J 2022 *Nature* **606** 890

[114]  Liu W, Cao L, Zhu S, Kong L, Wang G, Papaj M, Zhang P, Liu Y B, Chen H, Li G, Yang F, Kondo T, Du S, Cao G H, Shin S, Fu L, Yin Z, Gao H J, and Ding H 2020 *Nat. Commun.* **11** 5688

[115]  Xiong Y L, Guan J Q, Wang R F, Song C L, Ma X C, and Xue Q K 2022 *Chin. Phys. B* **31** 067401 (in Chinese)

[116]  Tan J H, Jiao N, Zheng M M, Zhang P, and Lu H Y 2025 *Chin. Phys. B* **34** 097402 (in Chinese)

[117]  Qu J Y, Hu G J, Xiang C L, Guo H, Lv S H, Han Y C, Xian G Y, Qi Q, Zhao Z, Zhu K, Lin X, Bao L H, Zou Y J, Sun L X, Yang H T, and Gao H J 2025 *Chin. Phys. B* **34** 067401 (in Chinese)

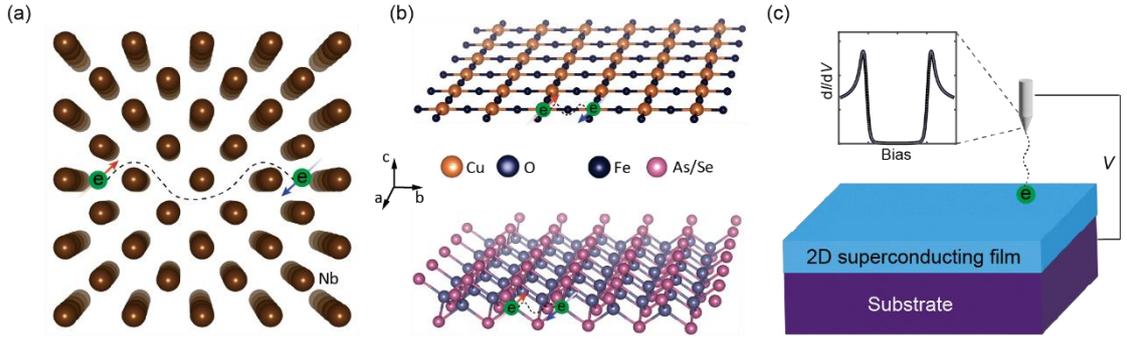

Fig. 1. From three- to two- dimensional superconductors. (a) Crystal of Nb superconductors where the electron pairing is long-range and isotropic. (b) Superconducting $CuO_2$ (top panel) and FeAs/Se (bottom panel) planes where the pairing is short-range and two-dimensional. (c) A sketch of microscopic spectroscopic measurement on superconductors.

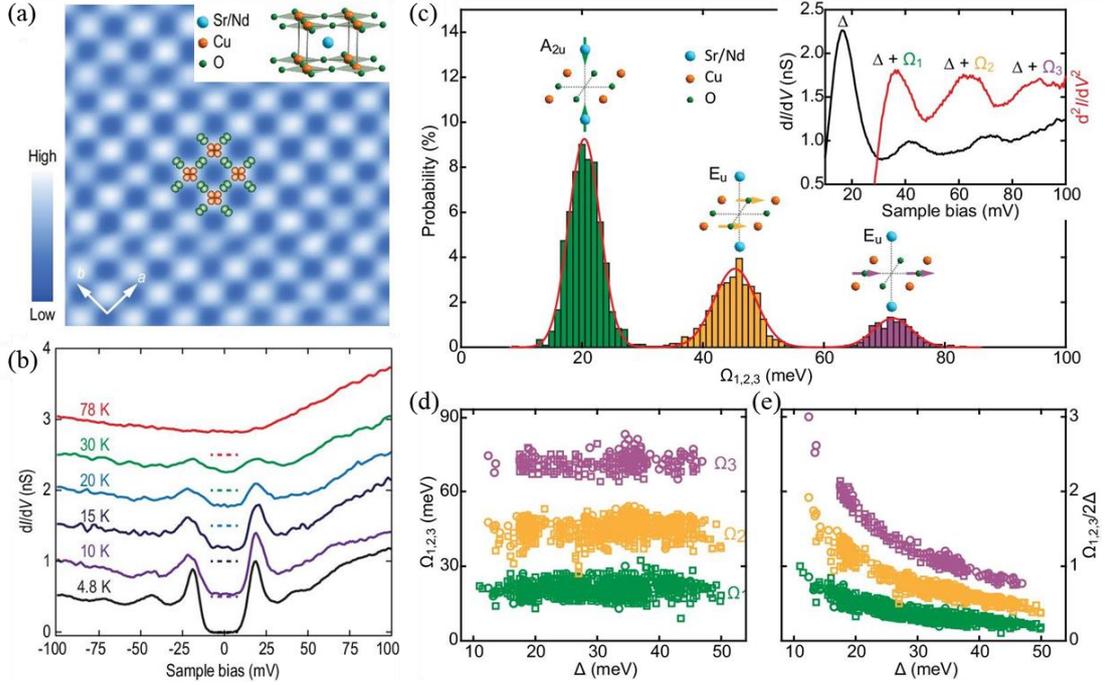

Fig. 2. Nodeless superconductivity on $CuO_2$ plane.[49] (a) Atomic-resolved image of $CuO_2$ terminated IL SNCO and its temperature evolution of U-shaped superconducting gap in (b). (c) The distribution of bosonic modes $\Omega_{1,2,3}$ and their relationship with gap size in (d). (e) The relationship between the ratio of $\Omega_{1,2,3}/2\Delta$ and gap size.

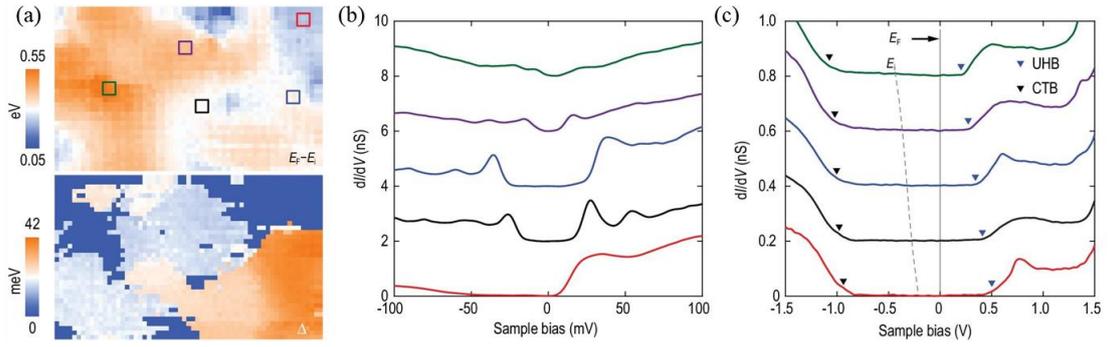

Fig. 3. Intrinsic inhomogeneity in $CuO_2$ plane.[49] (a) Mapping of CTG midpoint and gap size in the optimally doped SNCO. (b) Representative small-range STS collected in different local doping levels. (c) Large-range STS corresponding to the gap in (b).

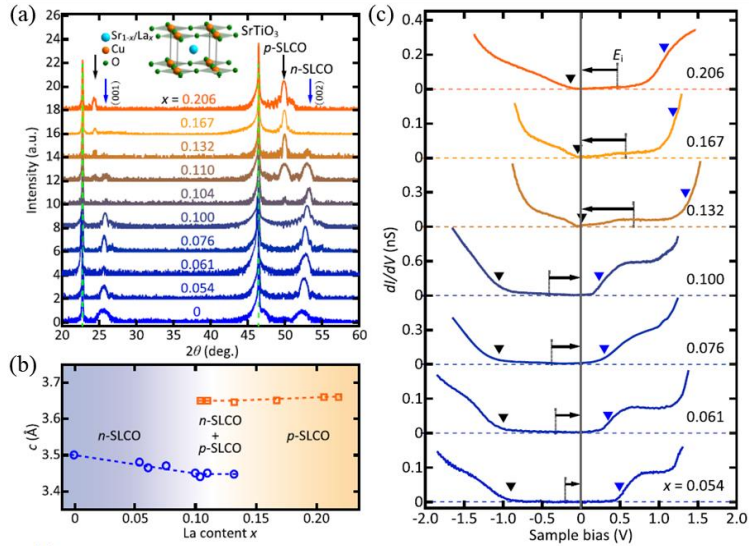

Fig. 4. Doping evolution of electronic structure in IL cuprates.[54] (a) Doping dependent XRD results of SLCO films. (b) Transition of $c$ axis length from n-SLCO to p-SLCO. (c) Systematic shift of CTG with increasing La content.

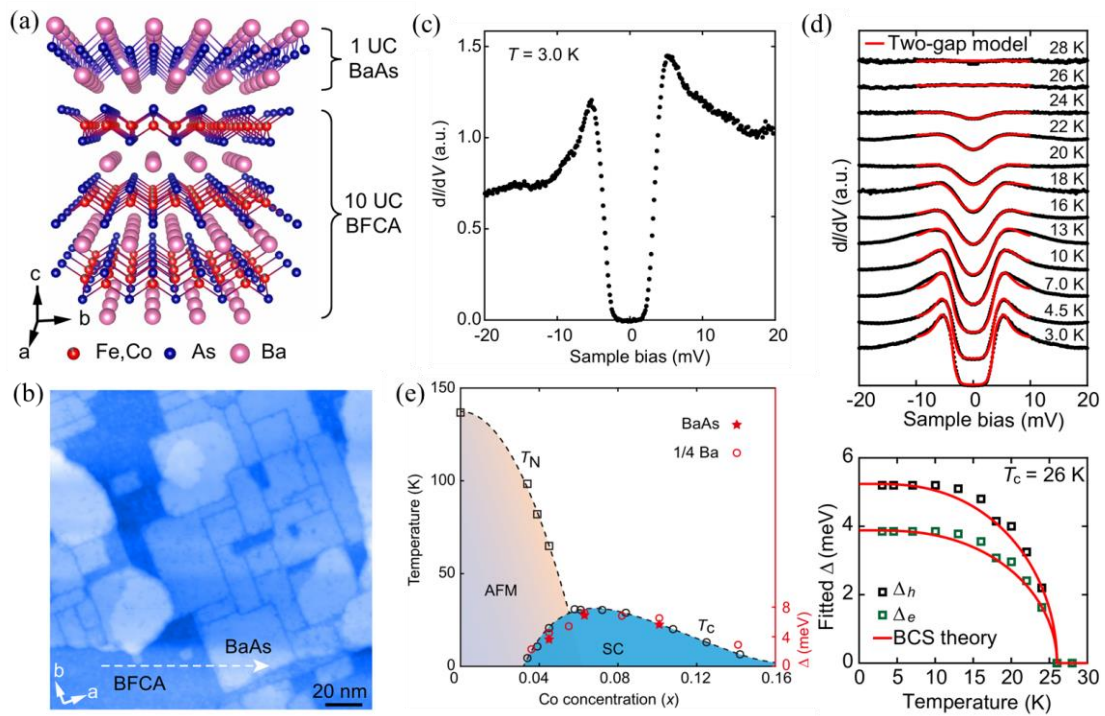

Fig. 5. Spectroscopic studies of FeAs plane.[64] (a) Structural model of BaAs/BFCA heterostructure. (b) STM topography of BaAs/BFCA heterostructure. (c) U-shaped gap on BaAs surface. (d) Temperature evolution of superconducting gap on BaAs surface. (e) Phase diagram of $T_c$ and superconducting gap of BaAs/BFCA.

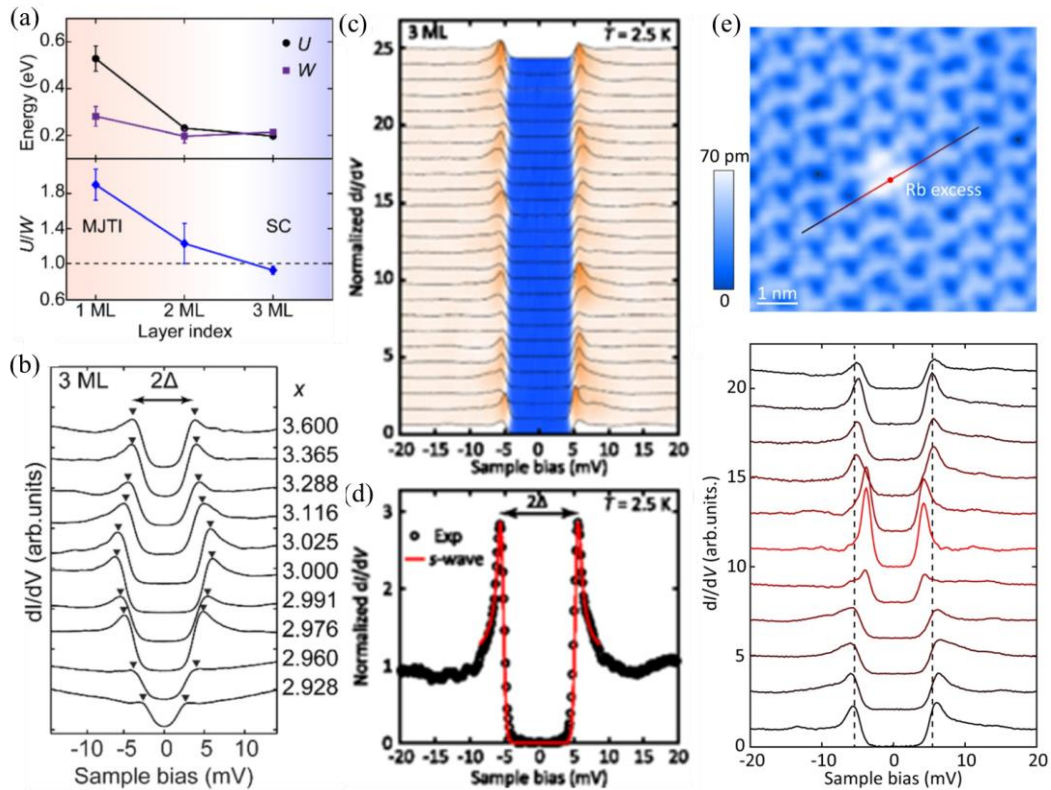

Fig. 6. Spectroscopic studies of $A_3C_{60}$. (a) Measured Hubbard U, W (upper panel), and $U/W$ (lower panel) of $K_3C_{60}$ as a function of thickness.[72] (b) Evolution of the averaged superconducting gap (measured at 4.5 K) with K doping $x$ in 3 ML.[72] (c) STS line cut and (d) d$I$/d$V$ spectrum taken at 2.5 K on 3 ML $K_3C_{60}$.[72] (e) STM topographies and d$I$/d$V$ spectra taken across a Rb excess.[73]

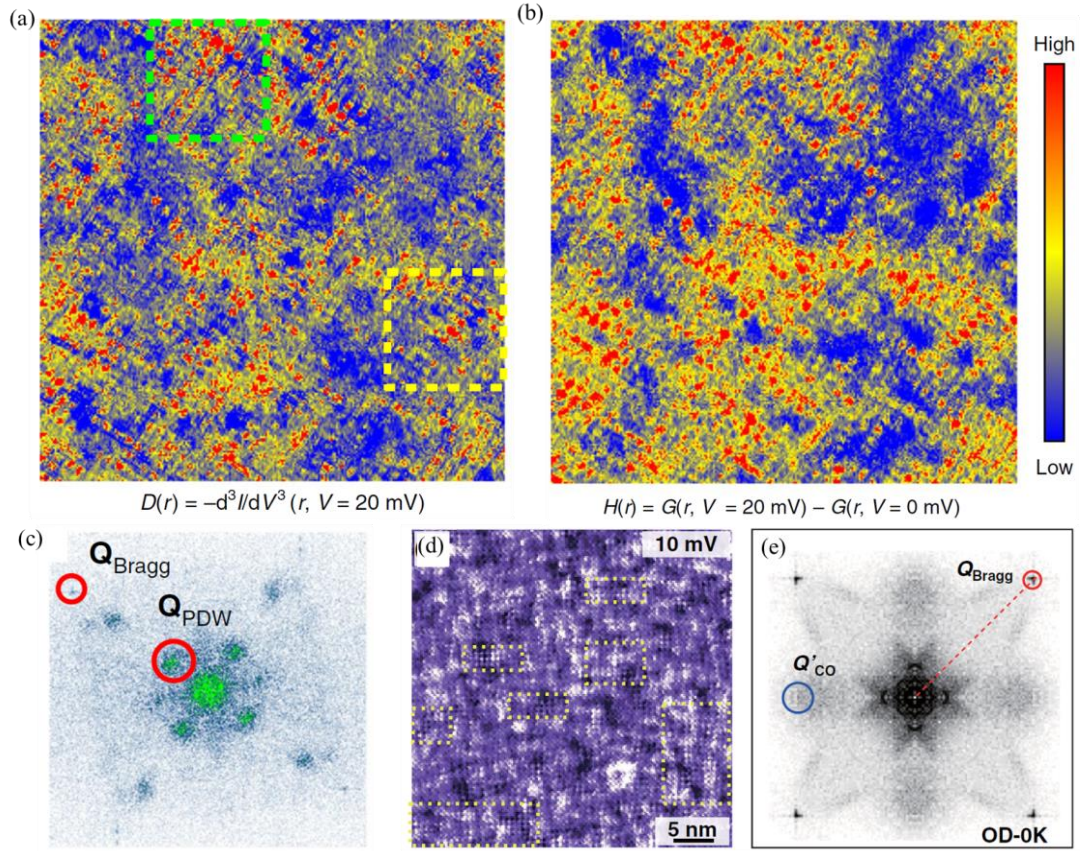

Fig. 7. PDW order in cuprates. (a) Spatial distribution of the superconducting coherence peak feature obtained from the D($r$) maps at V = 20 mV.[77] (b) The gap-depth map $H(r)$ defined as the spatially dependent difference of $G \equiv $ d$I$/d$V$ values at 20 mV and 0 mV.[77] (c) Fourier transform of the $H(r)$ map, with modulation wavevector $Q_{PDW} = (0.28 \pm 0.04)(2\pi/a_0)$.[77] (d) The d$I$/d$V$ map measured at bias voltage V = 10 mV of overdoped Bi-2201.[80] (e) The Fourier transform map of (d).[80]

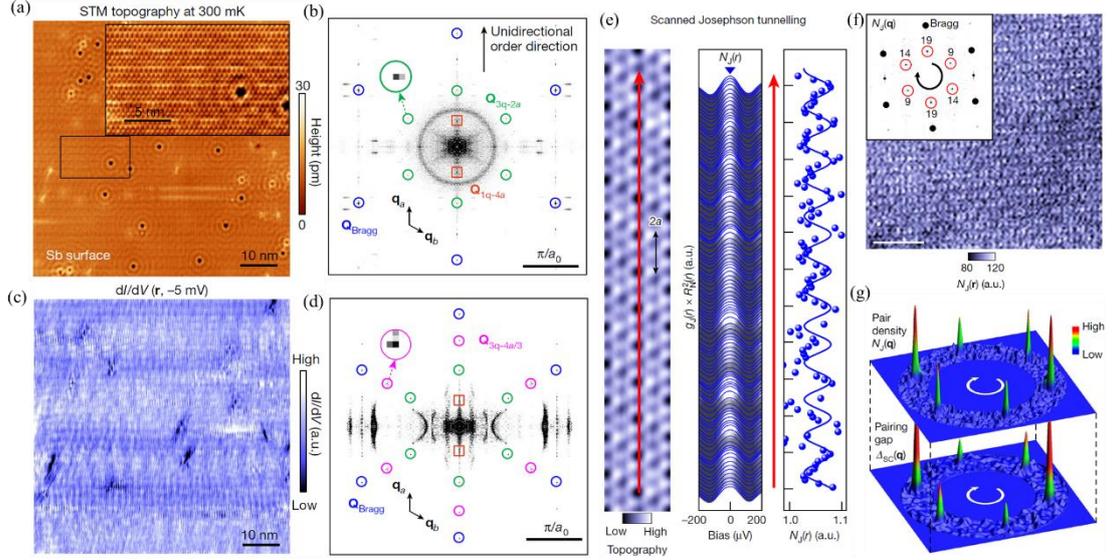

Fig. 8. PDW order in Kagome superconductors. (a) and (b) The large-scale STM topography of the Sb surface (a) and the magnitude of Fourier transform (b) in $CsV_3Sb_5$, showing 2 × 2 CDW and $4a_0$ unidirectional charge order.[81] (c) and (d) d$I$/d$V$($r$, −5 mV) map (c) and the magnitude of Fourier transform (d), revealing new PDW modulations.[81] (e) Atomically resolved Sb surface in $KV_3Sb_5$, differential conductance spectra $g_J(r, E)$ and the spatially resolved pair density $N_J(r)$ detected with the superconducting tip.[82] (f) Pair-density map of Sb surface detected by Josephson tunnelling microscopy.[82] (g) The 2 × 2 pairing-gap and pair-density modulations with chirality marked by the white arrow.[82]

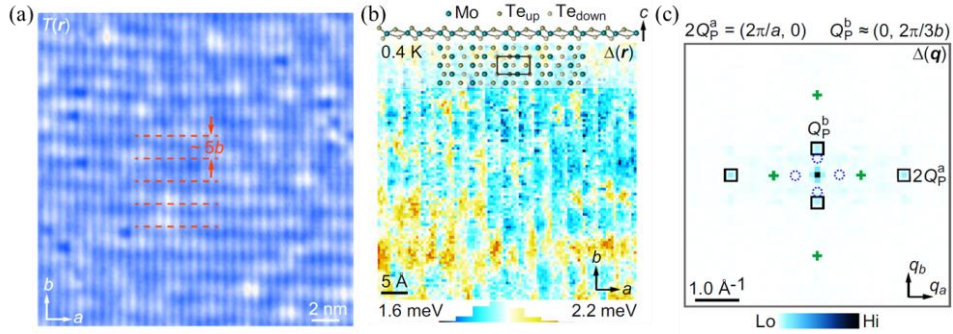

Fig. 9. PDW order in 1T'-$MoTe_2$. (a) Large-scale $T(r)$ of Te surface, presenting a unidirectional CDW along the $b$ axis at five unit-cell periodicity.[90] (b) Gap map |$\Delta(r)$| measured on the electron-doped 1T'-$MoTe_2$. Overlaid is a top view of the atomic structure.[91] (c) Derived |$\Delta(q)$| from |$\Delta(r)$| showing two sharp peaks corresponding to the PDW order.[91]

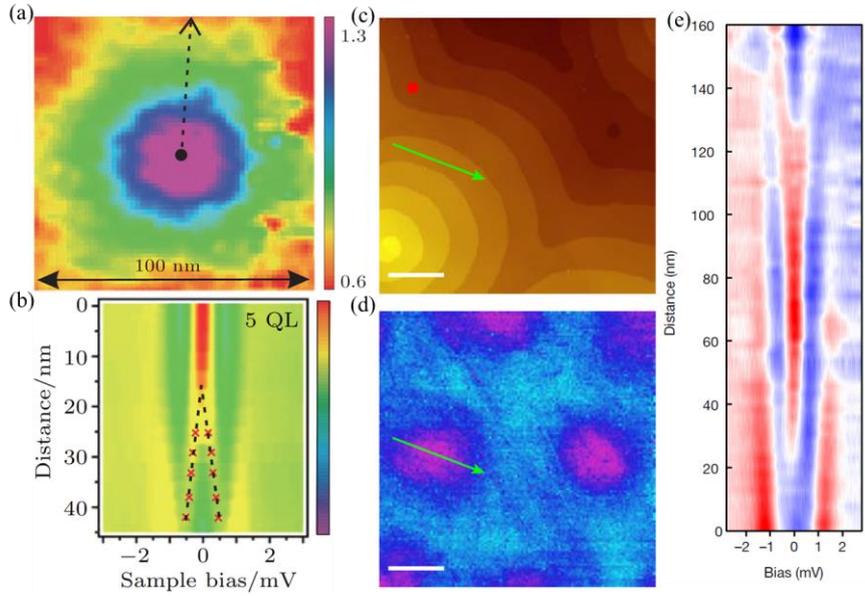

Fig. 10. MZMs in heterostructures. (a) A single Abrikosov vortex map and (b) Spectra across the vortex center of $Bi_2Te_3$/$NbSe_2$ heterostructure.[95] (c) STM image of approximately 80 nm SnTe(001) film grown on the 140 nm Pb film.[96] (d) Zero-bias $dI/dV$ map for the SnTe film at $T = 0.4$ K and $Bz = 0.02$ T.[96] (e) Spatially resolved $dI/dV$ spectra across the magnetic vortex.[96]

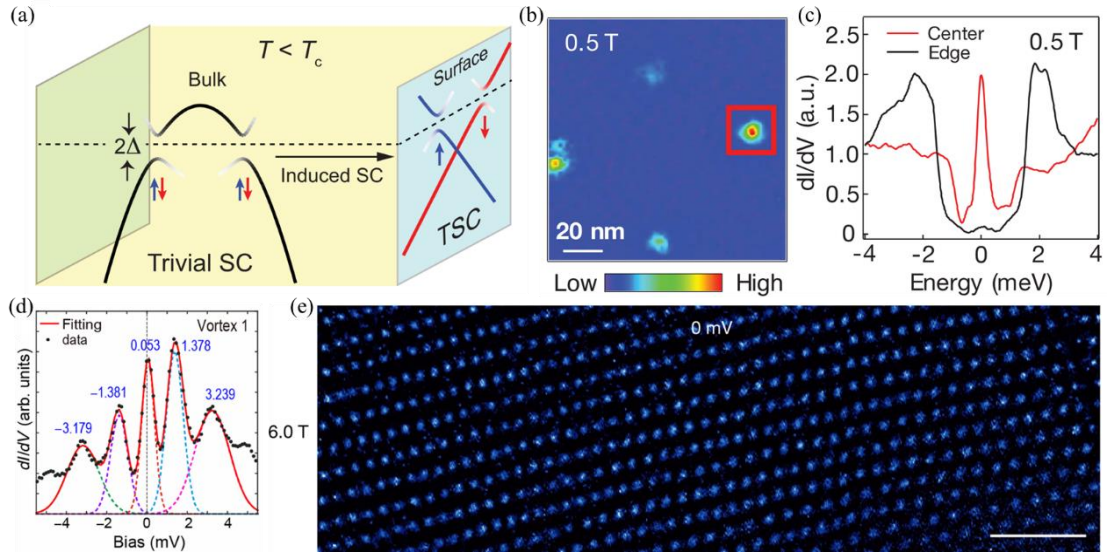

Fig. 11. MZMs in IBSCs. (a) Schematic of superconducting states in the bulk and on the surface of IBSCs.[105] (b) Zero-bias conductance (ZBC) map of $FeTe_{0.55}Se_{0.45}$ under 0.5 T magnetic field.[106] (c) $dI/dV$ spectra measured on the center and edge of the vortex in (b).[106] (d) Spatially resolved $dI/dV$ spectra in the magnetic vortex center of $(Li_{0.84}Fe_{0.16}OH)FeSe$.[107] (f) Micrometer-sized ordered MZM lattice under 6 T collected in strained LiFeAs.[111]